\documentclass{SCGE}

\renewcommand{\arraystretch}{2}

\usepackage{setspace}

\begin{document}

\begin{picture}(0,0){\rm
\put(0,-39){\makebox[160truemm][l]{\bf {\sanhao\raisebox{2pt}{.}}
Article  {\sanhao\raisebox{1.5pt}{.}}}}}
\put(0,-52){\jiuwuhao {\textcolor[rgb]{0.5,0.5,0.5}{\sf 
}}}
\end{picture}

\input psfig.sty
\def\bm{\boldsymbol}

\def\dl{\displaystyle}
\def\du{\end{document}}
\def\pi{{\uppi}}

\Year{2013} %
\Month{0} %
\Vol{0} 
\No{0} 
\BeginPage{0} 
\EndPage{0} 
\AuthorMark{{\rm Jianmin S}}  
\AuthorMarkCite{{\rm Jianmin S}.} 
\DOI{0} 
\title{Search for carbon stars and DZ white dwarfs in SDSS spectra survey through machine learning}

\author[1,2,3]{Jianmin Si}{}
\author[2,3]{Ali Luo\footnote{Corresponding author (email: lal@nao.cas.cn)}}{}
\author[2]{Yinbi Li}{}
\author[2]{Jiannan Zhang}{}
\author[2,3]{Peng Wei}{}
\author[1]{Yihong Wu}{}
\author[1]{\\Fuchao Wu}{}
\author[2]{Yongheng Zhao}{}

\address[1{\rm}]{National Laboratory of Pattern Recognition, Institute of Automation,
Chinese Academy of Sciences, 100190, Beijing, China}
\address[2{\rm}]{Key Laboratory of Optical Astronomy, National Astronomical Observatories, Chinese Academy of Sciences, Beijing, 100012, China} 
\address[3{\rm}]{University of Chinese Academy of Sciences, Beijing, 100049, China} 
\maketitle \vspace{-3.5mm}{\footnotesize\begin{center} Received October  ; accepted  ; published online  
\end{center}}\vspace*{-5mm}

\begin{center}
\rule{16.5cm}{0.4pt}
\parbox{16.5cm}
{\begin{abstract} 
\begin{spacing}{1.05}
Carbon stars and DZ white dwarfs are two types of rare objects in the Galaxy. In this paper, we have applied the label propagation algorithm to search for these two types of stars from Data Release Eight (DR8) of the Sloan Digital Sky Survey (SDSS), which is verified to be efficient by calculating precision and recall. From nearly two million spectra including stars, galaxies and QSOs, we have found 260 new carbon stars in which 96 stars have been identified as dwarfs and 7 identified as giants, and 11 composition spectrum systems (each of them consists of a white dwarf and a carbon star). Similarly, using the label propagation method, we have obtained 29 new DZ white dwarfs from SDSS DR8. Compared with PCA reconstructed spectra, the 29 findings are typical DZ white dwarfs. We have also investigated their proper motions by comparing them with proper motion distribution of 9,374 white dwarfs, and found that they satisfy the current observed white dwarfs by SDSS generally have large proper motions. In addition, we have estimated their effective temperatures by fitting the polynomial relationship between effective temperature and g-r color of known DZ white dwarfs, and found 12 of the 29 new DZ white dwarfs are cool, in which nine are between 6000K and 6600K, and three are below 6000K.
\end{spacing}
\end{abstract}}
\end{center}\vspace*{-0.6cm}

\begin{center}
\parbox{16.5cm}
{\bf\jiuhao  machine learning, label propagation, carbon stars, DZ white dwarfs,}

{}
\end{center}

\wuhao\vspace*{1.5mm}

\begin{multicols}{2}

\renewcommand{\baselinestretch}{1.08} \baselineskip 12.2pt\parindent=10.8pt

\renewcommand{\thefootnote}

\section{Introduction}\vspace*{-2mm}
\begin{spacing}{1.05}
\no The Sloan Digital Sky Survey (SDSS) is one of the most influential and successful sky surveys in the history of astronomy [1]. It uses a 2.5-meter telescope with two special-purpose instruments, a 120-megapixel camera and a pair of spectrographs fed by optical fibers. The camera can image 1.5 square degrees of sky one time, about eight times the area of the full moon, and the spectrographs can measure more than 600 spectra in a single observation. In addition, a set of software pipelines were designed to keep pace with the massive data flow. At present, SDSS-III has been publicly released raw and reduced data following SDSS-I and SDSS-II. The data of SDSS DR8 is available from http://www.sdss3.org/dr8/data\_access.php, following the earlier Data Release 1-7 of SDSS-I/II. Contained is all the imaging data over 14,000 square degrees of sky taken by the SDSS imaging camera, and also nearly two million spectra including 656,801 stars, 1,003,596 galaxies and 182,803 QSOs. 

From such massive spectra of SDSS, it is possible to find rare objects, although it is difficult to search.Machine learning is a suitable tool, and many machine learning algorithms have been widely used for finding unusual objects from massive spectra dataset. The algorithm based on local outlier factor (LOF) has been applied to search for supernovae from SDSS [2,3,4]. The Monte Carlo combined with LOF has been applied to detect outliers from SDSS DR8, and many rare objects were found including CV stars, binaries, emission line stars and one binary consisting of a supernova and a M2 type star [5]. The self-organizing mapping, which is one of neutral network algorithms, has been used to analyse QSOs at redshift from z=0.6 to 4.3 given by SDSS pipeline, and 1,005 peculiar QSOs were found [6]. The dimensionality reduction technique of local linear embedding (LLE), which preserves the local (and possibly non-linear) structure within high dimensional datasets, has been used to classify SDSS stellar spectra, and it was shown that the majority of stellar spectra could be represented as a one dimensional sequence within a three dimensional space. The spectra deviating from this sequence are spectra with emission lines (including misclassified galaxies) or broad absorption lines (e.g., carbon stars) [7]. The infinite Gaussian mixture model (GMM) has been applied to find variable star candidates in the Northern Sky Variability Survey [8].

Label propagation algorithm [9] is one of the semi-supervised ranking learning algorithms which has an important role in real world, because labeling samples is time consuming , whereas unlabeled samples are relatively accessible to obtain. The main principle  is to learn from both labeled and unlabeled samples, and similar samples have similar rankings under condition that labeled samples keep ranking top. Many different label propagation algorithms have been provided to date. Linear neighborhood propagation algorithm was proposed which assumes that each data point can be linearly reconstructed from neighbors, and it can be extended easily to out-of-sample data [10]. Sparsity induced similarity (SIS) measure was proposed which significantly improves label propagation performance [11]. A new framework named efficient manifold ranking (EMR) is a promising method to solve large scale retrieval problems, which significantly reduces the computational time [12]. Herein, we apply the label propagation algorithm, which supposes that a sample only has relationship to its k-nearest neighbors (KNN), to retrieve rare and interesting carbon stars and DZ white dwarfs from spectra of SDSS DR8.

Carbon stars are excellent halo dynamical tracers [13,14]. Battinelli et al. [15] extended rotation curve of the Milky Way to 24 kpc using radial velocity of carbon stars outside the Solar circle. It has been shown that the rotation curve displays a slight decline beyond the Solar circle. Demers et al. [16] determined the radial velocities for 70 carbon stars and the rotation velocities of carbon stars with $60^{\circ} < l < 150^{\circ}$ , and found a flat rotation curve of the Milky Way to 15 kpc. Demers et al. [17] used N-type carbon stars from the literature with measured radial velocity to investigate the thin disk rotation to unprecedented galactocentric distances, and concluded that the Milky Way rotation curve stays flat at least up to 20 kpc. Cool helium-rich white dwarfs showing traces of heavy elements (other than carbon) in their optical spectra are collectively known as DZ white dwarfs. They are rare, and more samples are required to analyse and understand the accretion problem in cool DZ white dwarfs. Aannestad et al. [18] derived radial velocities of 15 cool DZ white dwarfs using their H and K lines of Ca II, and concluded that only few stars could have just accreted or presently be accreting metals from local clouds. They also found that velocity of Ca accretion from local interstellar matter in the past could not have been sufficiently rapid to explain the photospheric abundances of Ca, at least in the cool metallic-line degenerates. Farihi et al. [19] analysed 146 DZ white dwarfs from the SDSS, and derived their calcium and hydrogen abundances, Galactic positions and kinematics. They concluded that there are no correlations found between their accreted calcium abundances and spatial-kinematical distributions relative to interstellar material.

There have been more than one thousand carbon stars and over one hundred DZ white dwarfs identified from SDSS dataset at present. Margon et al. [20] and Downes et al. [21] found 39 and 251 faint high Galaxy latitude carbon stars respectively using photometric methods, and Green [22] obtained 1220 high Galaxy latitude carbon stars using Cross-Corelation Function (CCF) from spectra dataset of SDSS DR8. Dufour et al. [23] identified 147 new DZ white dwarfs by visually inspecting spectra which were selected using photometric, and Koester et al. [24] identified 26 cool DZ white dwarfs which gapped low temperature sequence. However, carbons stars and DZ white dwarfs were not found completely, and it is significant to enlarge their catalogs especially using automatic method.

\section{The label propagation algorithm}\vspace*{-2mm}
\subsection{Problem setup}
Let $(x_{1}, \dots ,x_{l})$ be labeled samples and  $(x_{l+1}, \dots ,x_{l+u})$ be unlabeled samples. In general, labeled samples are very few compared with unlabeled samples, namely $l<<u$. To retrieve samples which have the same class as labeled samples from unlabeled samples is the problem to solve. It is unacceptable only using a few labeled samples to learn, and massive unlabeled samples should be considered. The principal of the label propagation algorithm is to learn from both labeled and unlabeled samples. Initially, we can assign labeled samples a high score, and then calculate the score of unlabeled samples subject to the constraint that similar samples have similar scores and the labeled samples keep high score. Finally, samples are ranked according to their score in descending order, and samples ranking top are results to search.
\subsection{Graph construction}  
The similarities between pairs of samples can be described well using a graph, because graph-based approaches offer promising performances via simple and intuitive graph representation. A graph can be interpreted using probability theory, such as probability transformation matrix, and model the manifold structures that may exist in massive data sources in high dimensional spaces, such as graph Laplacian matrix and local linear embedding (LLE) and so on. The graph $G=(X,W)$ consists of $n=l+u$ nodes, where $W$ is symmetric adjacency matrix, of which $W_{ij}$ represents similarity between sample $x_{i}$ and $x_{j}$. We constructed the adjacency matrix $W$ using k-nearest neighbors in Euclidean distance considering the following reasons. The constructed adjacency matrix should be sparse especially for massive dataset, because there are no similarities between many pairs of samples. There are usually two commonly accepted graph sparsification methods, KNN and $\epsilon$-neighborhood. Because the graph constructed using $\epsilon$-neighborhood is not often connected, we used KNN here. In addition, similarity metric is one of the most important and difficult problems, especially for high dimension dataset. The Euclidean distance widely used in machine learning is not considered a good metric method in global, however, it is robust in local. Thus the adjacency matrix is constructed using KNN in Euclidean distance. To improve the performance of the algorithm, many kernel methods can be used to make affine transformation. In this paper, we used Gaussian kernel method as follows:
\begin{equation}
S_{ij} = S(x_{i},x_{j}) = exp\left(-\frac{||x_{i}-x_{j}||^2}{c*\sigma_{ij}^2}\right)
\end{equation}
where $c$ is regularization factor. When $\sigma \to 0$, the result approaches propagation of 1NN, and when $\sigma \to \infty$ , all the weights are equal to 1. In this paper, we adopted $\sigma_{ij}^2 = mean(KNN(x_{i})*mean(KNN(x_{j}))$, namely average distance between $x_{i}$ and its k-nearest neighbors multiplies average distance between $x_{j}$ and its k-nearest neighbors. But because the adjacency matrix constructed using KNN is asymmetric, the symmetrization is made by following formula:

\begin{equation}
W_{ij} = max(S_{ij},S_{ji})
\end{equation}
\subsection{Label propagation algorithm description}
The algorithm can be described as follows:

\noindent Input:

\noindent $X=\{x_{1}, \dots ,x_{l},x_{l+1},\dots ,x_{l+u}\}$
 
\noindent $Y=\{y_{i}\}=\left\{ 
\begin{smallmatrix}
  1&  i=1,...,l \\
  0& \quad i=l+1,...l+u
\end{smallmatrix}
\right.
$
	
\noindent $F_{1}= \{f_{i}\} \in (0,1) \quad i=1, \dots ,l+u$ (initialized randomly)

\noindent $K:$number of returned samples ranking top	

\noindent 1) Construct the adjacency matrix $W$ as described in section 2.2.

\noindent 2) Construct matrix $P = D^{-1/2}WD^{-1/2}$, where D is a diagonal matrix with its $(i,i)$-element equals to the sum of the $i_{th}$ row of $W$.

\noindent 3) Iterate $F_{t+1} = \alpha PF_{t} + (1-\alpha )*Y$ until convergence, where $\alpha$ is a parameter in $(0,1)$.

\noindent 4) Sort $F = F_{t}$ in descending order and return index $Rank$ 

\noindent Output: 
 samples $x_{Rank_{1}},\dots ,x_{Rank_{K}}$ are our final results.

The step 3 can be inferred by minimizing the cost function $\mathcal{Q}(F)$ associated with $F$ as follows:
\begin{equation}
\mathcal{Q}(F) = 1/2(\sum_{i,j=1}^{n} W_{i,j}\parallel\frac{1}{\sqrt{D_{ii}}}F_{i} - \frac{1}{\sqrt{D_{jj}}}F_{j}\parallel^2 + \mu\sum_{i=1}^{n}\parallel F_{i} - Y_{i}\parallel^2)
\end{equation}
where $\mu >0$ is the regularization parameter.
The first term of the right side of the cost function is the smoothness constraint, which indicates that rankings of similar samples should keep similar. The second term is the fitting constraint, which means that rankings of label samples should remain at the top.The trade-off between the two terms can be tuned by the parameter $\mu$. The cost function can be solved using method of gradient descent.

Differentiating $\mathcal{Q}$ with respect to $F$, we have: 
\begin{equation}
\frac{\partial\mathcal{Q}}{\partial F}|_{F=F^*} = F^*-PF^* + \mu(F^*-Y) = 0
\end{equation}
which can be transformed into:
\begin{equation}
F^* = \frac{1}{1+\mu}PF^*+\frac{\mu}{1+\mu}Y
\end{equation}
let $\alpha = \frac{1}{1+\mu}$, we can have
$F_{t+1} = \alpha PF_{t} + (1-\alpha )*Y$ in step 3.

\section{Experiments and Results}\vspace*{-2mm}
\subsection{Feature selection}
A good feature, which describe data accurately, can improve the performance of an algorithm. For a spectrum, there are two important features to consider, spectral continuum and spectral lines. The continuum has close relationship with effective temperature, and spectral lines are closely related to many physical parameters including effective temperature, surface gravity and metal abundance. They are both contained in observed spectral flux. Because spectra line features can decrease the influence brought from bad flux calibration, and the method to calculate continuum may bring about errors, here, observed spectra and continuum-subtracted spectra are our selected features. In order to reduce the noise, median filter with width 5\AA \ is proceeded, which is efficient to remove the narrow skyline and noise. Continuum-subtracted spectra are obtained using the observed spectra to divide the pseudo-continuum which are obtained using median filter with width 300\AA. The two features are normalized by $L_{2}$-norm respectively using following formula:
\begin{equation}
f_{i} = f_{i}/\sqrt{\sum_{i=1}^{N} f_{i}^2} 
\end{equation}
where $f_{i}$ is flux for each wavelength and $N$ is the length of flux. 
Using the two features, we first construct two graph with adjacency matrix $W1$ and $W2$ using method in section 2.2, and then construct the final graph with adjacency matrix $W = W1 + W2$.
\subsection{Algorithm performance analysis}
Green [22] identified 1220 carbon stars from spectra of SDSS DR8 using cross-correlation function (CCF), which is the most complete carbon star catalog from SDSS at present. Using the cross identification tool of SDSS and positions of 1220 carbon stars with radius of two arcseconds $^1$\footnote{$^1$ http://skyserver.sdss3.org/dr8/en/tools/crossid/crossid.asp}, we totally obtained 1345 spectra for these stars, and a few of them were observed repeatedly . Among these spectra, 1313 were classified as star by SDSS pipeline, 10 as galaxy and 21 as QSO.

Supposing there are only 1313 carbon stars in 656,801 stellar samples, we were able to verify the efficiency and robustness of label propagation algorithm through searching for carbon stars from these samples by calculating recall and precision, which are two important evaluation methods for ranking learning algorithms. Recall is the fraction of relevant instances which are retrieved, while precision is the fraction of retrieved instances which are relevant. Here, recall and precision are able to calculate by $recall = n/1313$ and $precision=n/K$, where $K$ is returned sample numbers and $n$ is the number of carbon stars in the $K$ returned samples. We calculated the $recall$ and $precision$ for the cases of $K=2000$ and $K=3000$ using one of 20 labeled carbon star samples each time as described in Table 1, which were selected considering different types and qualities. For $K=2000$, all of the recalls are larger than 0.9 and 19 of them are larger than 0.96, and 19 precisions reach 0.6. For $K=3000$, although the precisions decrease to 0.42, 18 recalls are larger than 0.97, and two of them larger than 0.98. In addition, we found that 12 carbon stars in the samples of Green [22], which is listed in Table 2, are not included in the returned 3000 samples when we use all labeled samples of Table 1 to search for carbon stars. Through manually checking their spectra, we found their absences were primarily caused by poor spectral quality. The completeness of carbon stars identified by Green [22] were not checked, however, it is indeed shown that the label propagation algorithm is efficient.

\begin{table}[H]
\renewcommand{\arraystretch}{1.2}

\label{tab.table1}
\caption{Recalls and precisions of algorithm using different labeled samples}
\begin{center}\footnotesize \doublerulesep 0.2pt \tabcolsep 14pt
\begin{tabular}{p{2.0cm} p{1.3cm} p{1.3cm} p{1.3cm} p{1.3cm}}
\hline
labeled sample$^{\rm a)}$ &SNR$^{\rm b)}$ &recall,precision$^{\rm c)}$ & recall,precision\\
&&(K=2000)&(K=3000) \\ 
\hline  
1486-52993-368&4.38&0.9688,0.6360&0.9787,0.4283\\
1067-52616-602&5.61&0.9642,0.6330&0.9840,0.4307\\
2795-54563-618&5.89&0.9634,0.6325&0.9779,0.4280\\ 
453-51915-53&9.57&0.9634,0.6325&0.9749,0.4267\\ 
1311-52765-571&10.03&0.9627,0.6320&0.9787,0.4283\\ 
726-52207-239&6.10&0.9619,0.6315&0.9779,0.4280\\
1326-52764-502&5.10&0.9596,0.6300&0.9817,0.4297\\ 
1307-52999-116&23.42&0.9596,0.6300&0.9756,0.4270\\
1274-52995-359&4.24&0.9566,0.6280&0.9764,0.4273\\
1465-53082-570&21.27&0.9513,0.6245&0.9749,0.4267\\  
1687-53260-83&6.87&0.9505,0.6240&0.9741,0.4263\\   
1881-53261-165&7.72&0.9490,0.6230&0.9733,0.4260\\ 
696-52209-133&19.88&0.9490,0.6230&0.9733,0.4260\\
613-52345-344&5.66&0.9482,0.6225&0.9718,0.4253\\                       
1521-52945-596&8.35&0.9421,0.6185&0.9733,0.4260\\
2183-53536-447&36.47&0.9322,0.6120&0.9764,0.4273\\                       
2083-53359-93&4.75&0.9284,0.6095&0.9695,0.4243\\
2866-54478-351&71.65&0.9254,0.6075&0.9726,0.4257\\ 
3232-54882-307&37.53&0.9193,0.6035&0.9733,0.4260\\  
2619-54506-279&30.44&0.9018,0.5920&0.9695,0.4243 \\

\bottomrule[0.65pt] 
\end{tabular}
\end{center}

\vspace*{-1mm} {\footnotesize \hspace*{3.5mm}a) {spectra identifier using Plate-MJD-FiberID}

\hspace*{3.5mm}b) {SNR means median signal to noise ratio (SNR)}

\hspace*{3.5mm}c) {first number is recall and second is precision}
}
\end{table}

\begin{table}[H]
\renewcommand{\arraystretch}{1}
\label{tab:table2}
\caption{Stars from previous studies do not rank in the top 3000}
\begin{center}\footnotesize \doublerulesep 0.1pt \tabcolsep 5pt
\begin{tabular}{c c c}  %
\hline
Plate&MJD&FiberID\\ \hline   
1329&52767&421\\
1675&53466&47\\
1802&53885&202\\
2167&53889&336\\
2201&53904&296\\
2275&53709&270\\
2347&53757&582\\
2347&53757&602\\
2531&54572&585\\
2674&54097&453\\
724&52238&547\\
1084&52591&226\\
\bottomrule[0.65pt] 
\end{tabular}
\end{center}

\end{table}
\subsection{Search for carbon stars}
In order to search for carbon stars from spectra of SDSS DR8, we applied the label propagation algorithm on star, galaxy, QSO samples respectively, and selected the samples ranking top 5000 from star samples and top 500 samples from galaxy and QSO samples, respectively. As a result, there were 1,624 spectra identified as carbon stars including 1,584 spectra from star samples, 11 from galaxy samples and 29 from QSO samples. Removed repeated observation, the number of distinct carbon stars was 1,573. Finally, cross identification was made by positions of 1573 carbon stars with database of SIMBAD, NED and ADS, and found that 260 carbon stars listed in Table 3 had not been previously reported. 
 	
In order to identify dwarf and giant carbon stars, we used the criteria described in Green [22]. Dwarf carbon stars should have significant proper motions which satisfy conditions as follows: (1) at least one USNO-B detection and one SDSS detection per source ($nfit>2$); (2) the proper motion in at least one coordinate larger than $3 \sigma$ , where $\sigma$ is the proper motion uncertainty in that coordinate; (3) total proper motion larger than 11 mas/yr. Giant carbon stars should satisfy conditions as follows: (1) have no significant proper motion; (2) i-band magnitude should be lower than 16.4. In our carbon stars, 96 were identified as dwarfs, and 7 are giants. 

In addition, 11 rare and significant composite spectrum systems consisting of a white dwarf and a carbon star were found in Table 4, three of which were observed twice.We decomposed them into a white dwarf and a carbon star using least square fitting. The white dwarf components are from white dwarf samples identified by Kleinman et al. [25] with SNR larger than 15, and carbon star components are 10 PCA components obtained by carbon samples found by Green [22] with SNR larger than 15. The results are described in Fig. 1. The black curve is composition spectrum, the red curve is white dwarf component, and the green curve is reside spectrum obtained using composition spectrum to subtract white dwarf component. Ten of 11 white dwarf components are DA white dwarfs, and one is PG1159 star also called  pre-degenerate, a star with a hydrogen-deficient atmosphere which is in transition between being the central star of a planetary nebula and being a hot white dwarf. The 10 composition spectra with DA white dwarf components are possibly rare and significant DA/dC  composite systems which are the definitive samples that strongly support the post binary mass transfer scenario for the origin of dCs. So far only two DA/dC composite systems have been identified and reported to date. The first one PG0824+289 was identified by Herber et al. [26], and the second one SBS 1517+5017 was reported by Liebert et al. [27]. In addition, nine of such systems were found by Green [22], but they were not listed which will be discussed elsewhere.Therefore, we guess a portion of our composition spectra are DA/dC systems found by Green [22]. 
\end{spacing}
\end{multicols}

\begin{table}[H]

\label{tab:table1}
\caption{New carbon star catalog}
\renewcommand{\arraystretch}{1.1}
\begin{center}\footnotesize \doublerulesep 0.2pt \tabcolsep 3pt
\begin{tabular*}{\textwidth}{l *{13}c}
\hline
Catalog&\multicolumn{1}{c}{u}&\multicolumn{1}{c}{g}&\multicolumn{1}{c}{r}&\multicolumn{1}{c}{i}&\multicolumn{1}{c}{z}&\multicolumn{1}{c}{j}&\multicolumn{1}{c}{h}&\multicolumn{1}{c}{k}&\multicolumn{1}{c}{$\mu_{\alpha} cos(\delta )$} & \multicolumn{1}{c}{$\mu_{\delta}$}&\multicolumn{1}{c}{$\mu_{\alpha} cos(\delta )\_ err$}&\multicolumn{1}{c}{$\mu_{\delta}\_ err$}&\multicolumn{1}{c}{class$^{\rm a)}$}\\
 &\multicolumn{1}{c}{(mag)}&\multicolumn{1}{c}{(mag)}&\multicolumn{1}{c}{(mag)}&\multicolumn{1}{c}{(mag)}&\multicolumn{1}{c}{(mag)}&\multicolumn{1}{c}{(mag)}&\multicolumn{1}{c}{(mag)}&\multicolumn{1}{c}{(mag)}&\multicolumn{1}{c}{($mas yr^{-1}$)}& \multicolumn{1}{c}{($mas yr^{-1}$)}&\multicolumn{1}{c}{($mas yr^{-1}$)}&\multicolumn{1}{c}{($mas yr^{-1}$)}&\\
\hline
SDSS J000023.28+560908.2&NA$^{\rm b)}$&NA&NA&NA&NA&NA&NA&NA&NA&NA&NA&NA&u\\
SDSS J000207.44+563755.1&NA&NA&NA&NA&NA&NA&NA&NA&NA&NA&NA&NA&u\\
SDSS J000259.39-050155.2&17.56&16.34&15.83&15.68&15.63&NA&NA&NA&-19.94&-46.69&2.4&2.4&d\\
SDSS J000341.08+235537.0&22.04&19.24&17.33&16.63&16.07&NA&NA&NA&18.3&3.39&2.93&2.93&d\\
SDSS J002727.61+063935.0&18.88&17.99&17.54&17.4&17.32&16.49&16.23&16.23&14.5&-15.29&2.67&2.67&d\\
SDSS J003118.08-092857.7&19.48&18.09&17.46&17.28&17.16&NA&NA&NA&-4.99&-4.15&3.1&3.1&u\\
SDSS J003537.50-181514.5&17.1&16.16&15.73&15.55&15.44&NA&NA&NA&53.91&-21.78&2.32&2.32&d\\
SDSS J003608.80+145823.0&20.01&18.67&17.88&17.62&17.48&16.58&16.22&16.01&2.85&-5.52&3.23&3.23&u\\
SDSS J003701.42+251825.0&18.3&16.73&16.09&15.9&15.85&NA&NA&NA&14.01&9.16&2.29&2.29&d\\
SDSS J003954.70+135604.7&18.64&16.78&15.98&15.69&15.56&14.56&14.11&14.03&6.3&-0.43&2.73&2.73&u\\
SDSS J004801.93+000351.8&19.92&18.82&18.25&18.06&18.01&17.11&16.77&16.7&26.53&-9.17&3.01&3.01&d\\
SDSS J004926.14-004340.8&17.91&16.59&15.93&15.73&15.64&14.7&14.33&14.3&0.39&-9.51&2.53&2.53&g\\
SDSS J005253.14+004841.1&22.52&20.73&19.13&18.29&17.83&16.39&15.75&15.28&54.92&-13.51&4.27&4.27&d\\
SDSS J010049.43-010111.5&24.63&25.11&19.59&17.81&22.83&NA&NA&NA&NA&NA&NA&NA&u\\
SDSS J010511.16-095852.3&19.89&18.77&18.25&18.04&17.95&NA&NA&NA&21.97&-2.26&3.29&3.29&d\\
SDSS J010549.68-103013.7&19.15&18.13&17.63&17.48&17.42&NA&NA&NA&24.51&-11.22&3.04&3.04&d\\
SDSS J011535.66-103806.3&19.81&18.36&17.8&17.6&17.52&NA&NA&NA&2.27&-1.84&2.97&2.97&u\\
SDSS J015227.77+125357.5&19.3&17.75&17.09&16.82&16.65&15.72&15.26&15.21&6.33&0.85&2.69&2.69&u\\

\bottomrule[0.65pt] 
\end{tabular*}
\end{center}
\vspace*{-1mm} {\footnotesize \hspace*{10.5mm}a) {'d' indicates identified dwarf carbon star; 'g' indicates identified giant carbon star; 'u' indicates uncertain}

\hspace*{10.5mm}b) {'NA' indicates corresponding parameters, for example, magnitude or proper motion, are not available}

\hspace*{10.5mm} {(Entire Table 3 can be found from the online journal. A portion shown here is for guidance regarding form and content)}}
\end{table}

\begin{table*}
\label{tab:table1}
\caption{Composition spectrum systems}
\renewcommand{\arraystretch}{1.1}
\begin{center}\footnotesize \doublerulesep 0.2pt \tabcolsep 15pt
\begin{tabular*}{\textwidth}{l *{10}c}
\hline
Catalog&Plate&MJD&FiberID&\multicolumn{1}{c}{u}&\multicolumn{1}{c}{g}&\multicolumn{1}{c}{i}&\multicolumn{1}{c}{r}&\multicolumn{1}{c}{z}\\
&&&&(mag)&(mag)&(mag)&(mag)&(mag)\\
\hline
SDSS J013007.13+002635.3$^{\rm a)}$&697&52226&468&19.29&18.75&17.73&17.28&16.96\\
&400&51820&342&\multicolumn{1}{c}{--$^{\rm b)}$}&\multicolumn{1}{c}{--} &\multicolumn{1}{c}{--}&\multicolumn{1}{c}{--}&\multicolumn{1}{c}{--}\\
SDSS J080908.15+360128.9&892&52378&431&19.54&19.33&18.85&18.66&18.53\\
SDSS J081157.14+143533.0&2270&53714&253&15.74&15.97&15.74&15.56&15.38\\
SDSS J090302.86+385527.5$^{\rm a)}$&936&52705&536&17.89&18.18&17.87&17.51&17.29\\
&3149&54806&436&\multicolumn{1}{c}{--}&\multicolumn{1}{c}{--}&\multicolumn{1}{c}{--}&\multicolumn{1}{c}{--}&\multicolumn{1}{c}{--}\\
SDSS J101548.90+094649.8&1597&52999&135&18.42&17.81&16.84&16.46&16.22\\
SDSS J103513.57+394743.4&1432&53003&434&19.7&19.41&18.67&18.19&17.95\\
SDSS J103837.22+015058.5&505&52317&107&17.25&17.48&17.55&17.41&17.29\\
SDSS J110747.05+154219.6&2491&53855&222&19.36&18.91&17.93&17.47&17.1\\
SDSS J112813.94+125234.3&1606&53055&460&19.38&18.95&18.32&17.86&17.52\\
SDSS J144910.15+301853.9&1843&53816&300&20.14&19.96&19.36&18.96&18.76\\
SDSS J151905.96+500702.9$^{\rm a)}$&1166&52751&275&17.4&17.57&17.42&17.05&16.87\\
&3297&54941&53&\multicolumn{1}{c}{--}&\multicolumn{1}{c}{--}&\multicolumn{1}{c}{--}&\multicolumn{1}{c}{--}&\multicolumn{1}{c}{--}\\

\bottomrule[0.65pt] 
\end{tabular*}
\end{center}  
\vspace*{-1mm} {\footnotesize \hspace*{10.5mm}a) {objects are observed twice}}
\hspace*{10.5mm}¡¡{b) '--' indicates same magnitude as above row}
\end{table*}

\begin{figure*}
\begin{center}
\includegraphics[width=0.9\textwidth]{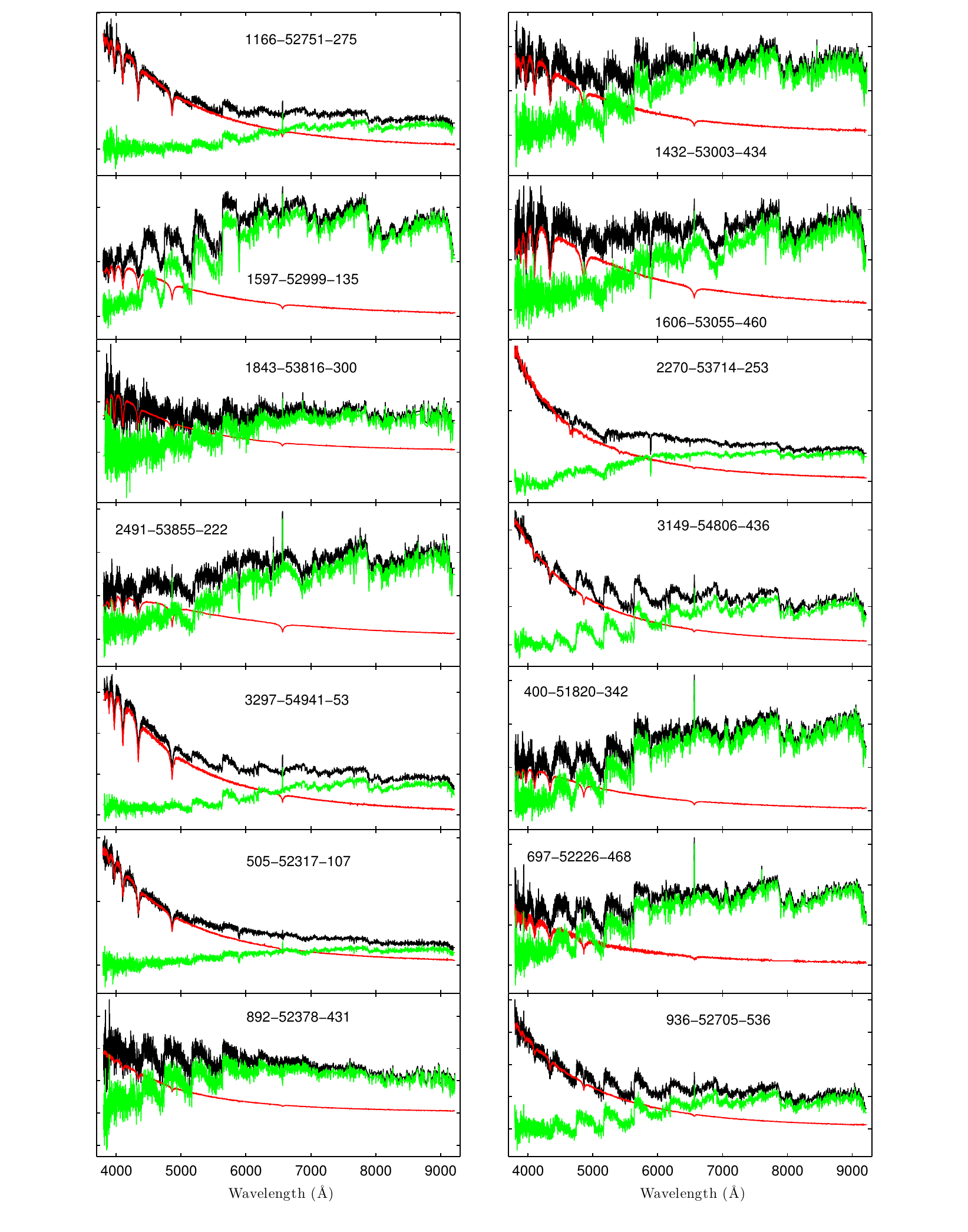} \\ 
\end{center}
\label{RL4}
\caption{Decomposition of the composition spectrum systems. Black curve is the observed composition spectrum system, red is the decomposed white dwarf component and green is the decomposed carbon star component. Number of each sub-figure are the Plate, MJD and FiberID of the spectrum.}
\end{figure*}

\begin{figure*}
\begin{center}
\includegraphics[width=0.9\textwidth]{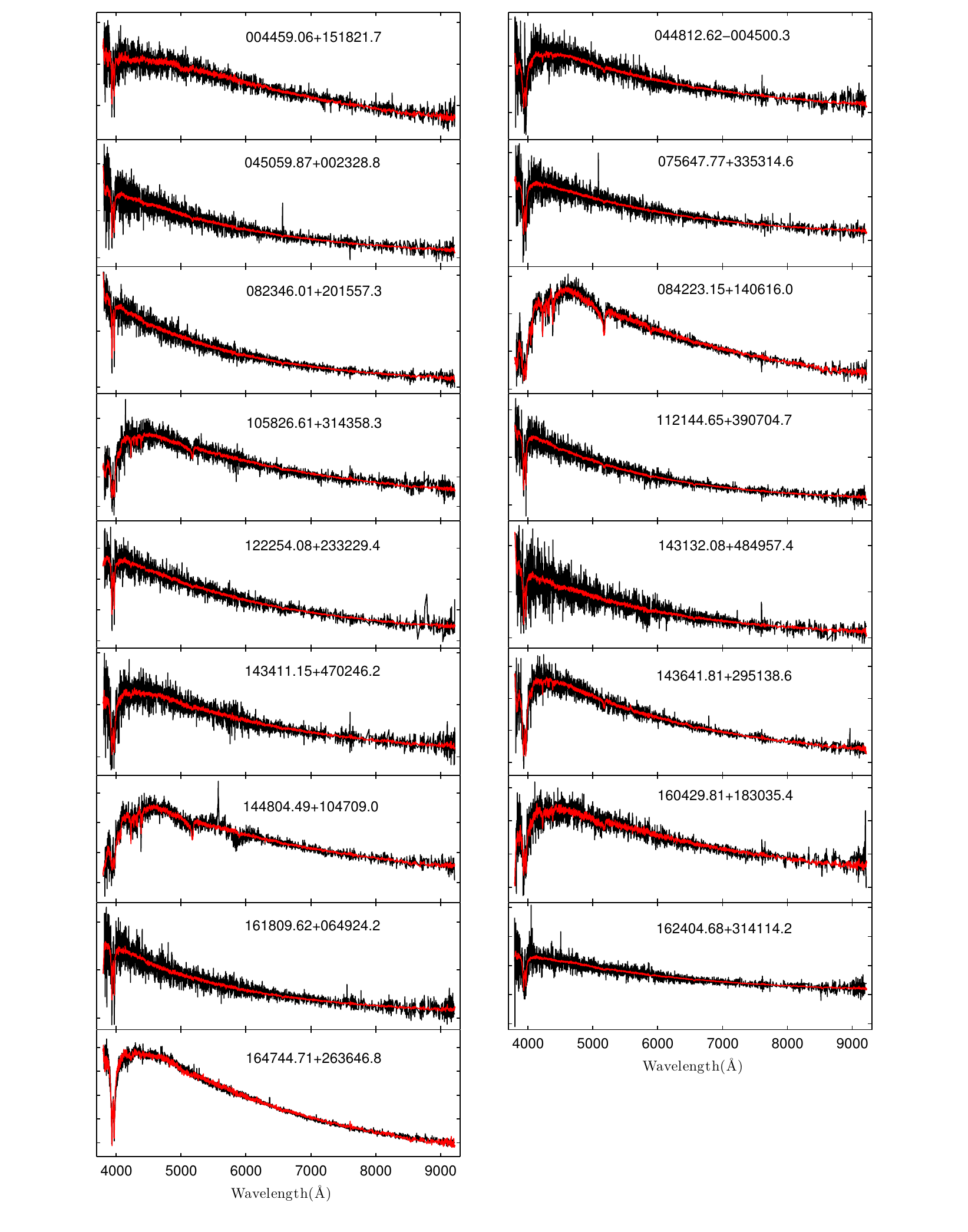} \\ 
\end{center}
\label{RL4}
\caption{DZ white dwarfs reconstruction with temperature larger than 6600K. Black curve is the observed spectra and red is the reconstructed spectra using PCA algorithm}
\end{figure*}

\begin{figure*}
\begin{center}
\includegraphics[width=0.9\textwidth]{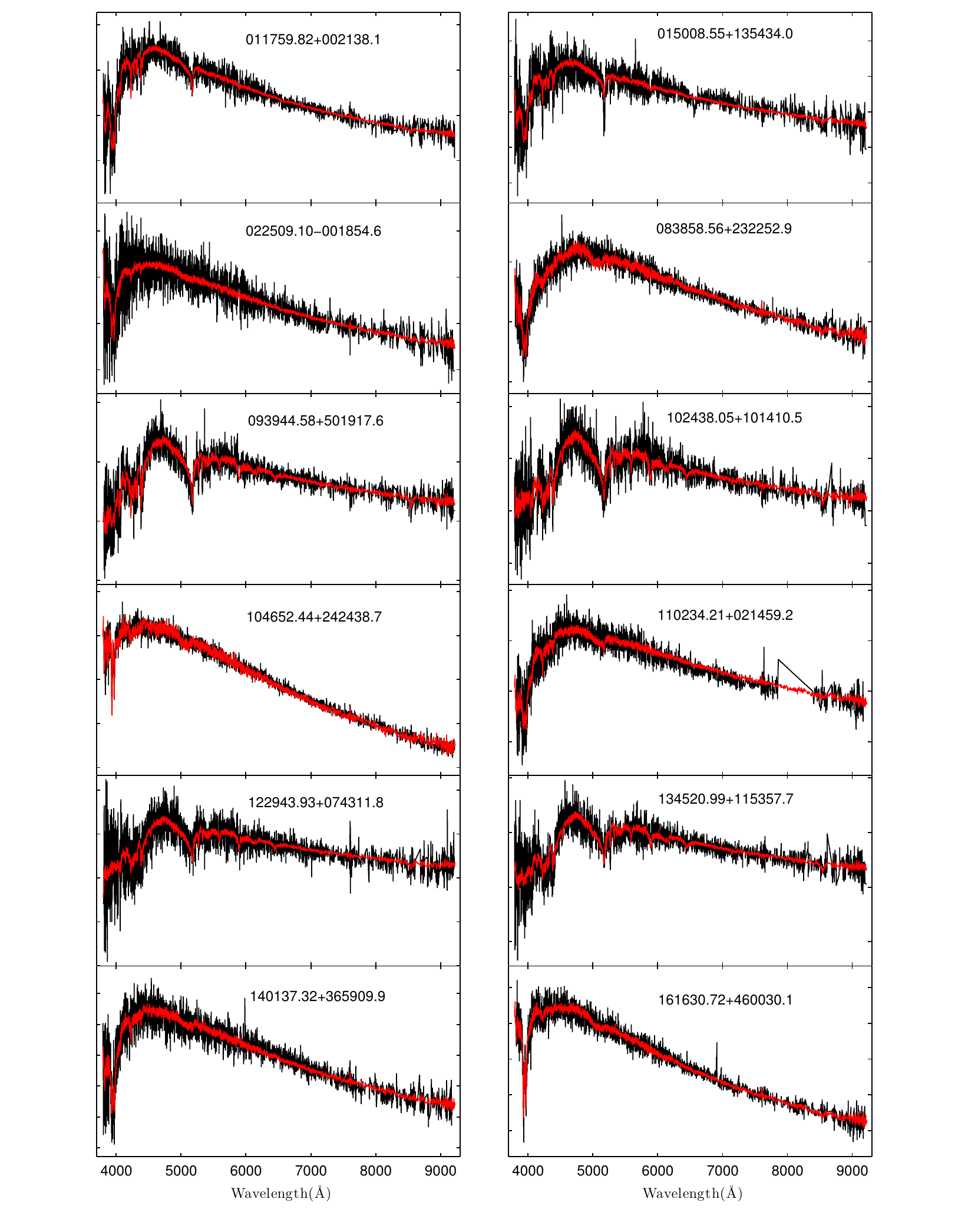} 
\end{center}
\label{RL4}
\caption{DZ white dwarfs reconstruction with temperatures below 6600K. Black curve is the observed spectra and red is the reconstructed spectra using PCA algorithm}
\end{figure*}

\begin{multicols}{2}
\subsection{Search for DZ white dwarfs}
\begin{spacing}{1.05}
In the same way, we searched for DZ white dwarfs from star, galaxy and QSO samples of SDSS DR8 respectively using label propagation algorithm. We selected the samples ranking top 1000 from star and galaxy samples respectively, and 2000 from QSO samples. A total of 222 DZ white dwarf spectra were found, including 29 spectra from star samples, 8 from galaxy samples and 185 from QSO samples. The number of distinct DZ white dwarfs is 188 because of repeated observation. Then we made a cross identification with database of SIMBAD, NED and ADS, and found that 29 stars had not been reported elsewhere and are listed in Table 5. 

\quad We reconstructed our new 29 DZ white dwarfs by PCA algorithm using 173 DZ white dwarfs identified by Dufour et al. [23] and Koester et al. [24]. First, these 173 spectra were denoised using Gaussian filter with FWHM= 3 \AA \ . Then, the PCA algorithm was applied to these denoised spectra, and 10 principle components obtained. After these steps, our samples were reconstructed using these 10 principle components. If they are typical DZ white dwarfs, they should be reconstructed well using the 10 principle components. The constructed results are illustrated in Fig. 2 and Fig. 3, and it can be noted that the 29 stars are reconstructed well particularly for the region of CaII H \& K lines or MgI line. 

\quad In addition, we used the white dwarf catalog of Kleinman [25] from SDSS with number of 19,712 to make cross-identification with the proper motion catalog of Munn et al. [28], which combines astrometry from the USNO-B and SDSS imaging surveys, and obtained  proper motions of 13,592 white dwarfs, among which proper motions of 9,374 white dwarfs have high confidence levels. We plotted proper motion accumulation distribution of these 9,374 white dwarfs in Fig. 4, and can conclude that the white dwarfs currently observed by SDSS generally have large proper motions, and proper motions of 99\% white dwarfs are larger than 10.5 mas/yr. In order to investigate whether our new DZ white dwarfs have large proper motions, we compared them with the accumulation distribution. Proper motions of our 29 DZ white dwarfs were obtained by making cross-identification with different catalogs and are listed in Table 5, in which 26 are from catalog of Munn et al. [28], one is from PPMXL catalog, and two are not available. In addition, two DZ white dwarfs, i.e., SDSS J045059.87+002328.8 and  SDSS J075647.77+335314.6, have proper motions with low confidence levels. We marked proper motions of our 25 samples with high confidence levels as asterisk in the accumulation distribution of Fig. 4, and found that proper motions of our 21 samples were larger than 65\% of 9,374 samples. Our minimum proper motion is 18.2 mas/yr, larger than 20\% of 9,374 samples. Thus, the 25 DZ white dwarfs do satisfy the fact that the white dwarfs currently observed by SDSS have large proper motions in general.
\end{spacing}
\end{multicols}
\begin{table}

\label{tab:table6}
\caption{New DZ white dwarfs}
\renewcommand{\arraystretch}{1.1}
\begin{center}\footnotesize \doublerulesep 0.2pt \tabcolsep 5pt
\begin{tabular*}{\textwidth}{l *{12}c}
\hline
Catalog&\multicolumn{1}{c}{u}&\multicolumn{1}{c}{g}&\multicolumn{1}{c}{r}&\multicolumn{1}{c}{i}&\multicolumn{1}{c}{z}&\multicolumn{1}{c}{$T\_{eff}$}&\multicolumn{1}{c}{$\mu_{\alpha} cos(\delta )$} & \multicolumn{1}{c}{$\mu_{\delta}$}&\multicolumn{1}{c}{$\mu_{\alpha} cos(\delta )\_ err$}&\multicolumn{1}{c}{$\mu_{\delta}\_ err$}&\multicolumn{1}{c}{$\mu$}\\
&(mag)&(mag)&(mag)&(mag)&(mag)&(K)&($mas yr^{-1}$)& ($mas yr^{-1}$)&($mas yr^{-1}$)&($mas yr^{-1}$)&($mas yr^{-1}$)&\\
\hline
SDSS J004459.06+151821.7$^{\rm a)}$&20.15&19.44&19.14&19.05&19.11&6283&60.3&17.1&5.6&5.6&62.68\\
SDSS J011759.82+002138.1&20.19&19.32&19.12&19.17&19.37&6732&NA$^{\rm b)}$&NA&NA&NA&NA\\
SDSS J015008.55+135434.0&21.03&20.15&19.92&19.89&19.85&6581&81.62&-51.82&3.77&3.77&96.68\\
SDSS J022509.10-001854.6&20.97&20.05&19.81&19.82&19.84&6534&22.1&-43.65&3.79&3.79&48.93\\
SDSS J044812.62-004500.3&20.53&20.26&20.27&20.47&20.49&8320&14.52&-17.67&5.81&5.81&22.87\\
SDSS J045059.87+002328.8&20.49&20.16&20.20&20.27&20.35&8649&5.44&-9.09&5.97&5.97&10.59\\
SDSS J075647.77+335314.6&20.37&19.96&19.95&20.01&20.12&8118&-6.13&0.24&4.86&4.86&6.14\\
SDSS J082346.01+201557.3&18.96&18.82&18.95&19.16&19.30&9856&-29.14&-27.11&3.23&3.23&39.8\\
SDSS J083858.56+232252.9&20.19&19.21&18.86&18.82&18.83&6109&33.59&-24.81&3.3&3.3&41.76\\
SDSS J084223.15+140616.0&19.29&18.31&18.17&18.27&18.44&7082&NA&NA&NA&NA&NA\\
SDSS J093944.58+501917.6&23.00&19.95&19.48&19.50&19.68&5797&-2.93&-81.38&5.5&5.5&81.43\\
SDSS J102438.05+101410.5&23.06&20.59&20.03&20.13&20.19&5639&-51.02&-4.5&4.82&4.82&51.22\\
SDSS J104652.44+242438.7&18.71&18.05&17.81&17.78&17.79&6534&-67.89&-72.5&2.84&2.84&99.33\\
SDSS J105826.61+314358.3&19.81&19.09&18.95&19.01&19.18&7082&16.93&-34.61&4.51&4.51&38.53\\
SDSS J110234.21+021459.2&21.22&19.53&19.20&19.17&19.30&6175&-93.88&21.82&3.63&3.63&96.38\\
SDSS J112144.65+390704.7&19.09&18.93&19.10&19.30&19.60&10522&-63.76&15.08&3.34&3.34&65.52\\
SDSS J122254.08+233229.4&19.27&19.03&19.10&19.15&19.25&9012&-64.4&-37.23&2.93&2.93&74.38\\
SDSS J122943.93+074311.8&23.31&20.42&20.07&20.07&20.33&6109&18.2&-0.17&5.01&5.01&18.2\\
SDSS J134520.99+115357.7&22.64&19.92&19.39&19.45&19.54&5686&-118.85&73.75&3.71&3.71&139.87\\
SDSS J140137.32+365909.9&19.94&18.81&18.53&18.52&18.55&6361&-24.33&-54.61&2.91&2.91&59.79\\
SDSS J143132.08+484957.4&20.19&19.95&20.05&20.23&20.26&9413&-33.28&16.39&3.72&3.72&37.09\\
SDSS J143411.15+470246.2&20.21&19.79&19.81&19.90&20.10&8426&-52.48&50.21&3.87&3.87&72.63\\
SDSS J143641.81+295138.6&19.31&18.67&18.67&18.87&19.04&8217&2.49&-18.04&2.96&2.96&18.21\\
SDSS J144804.49+104709.0&20.18&18.81&18.65&18.67&18.75&6957&45.78&-44.46&3.21&3.21&63.81\\
SDSS J160429.81+183035.4&20.70&19.70&19.55&19.61&19.76&7019&-19.66&-51.43&3.16&3.16&55.06\\
SDSS J161630.72+460030.1&19.48&18.76&18.51&18.49&18.59&6489&-160.45&53.26&3.34&3.34&169.06\\
SDSS J161809.62+064924.2&20.07&19.91&19.99&20.14&20.21&9141&16.69&-60.72&5.15&5.15&62.97\\
SDSS J162404.68+314114.2&19.87&19.45&19.61&19.70&19.88&10347&22.51&5.73&5.52&5.52&23.23\\
SDSS J164744.71+263646.8&17.53&17.00&16.92&16.96&17.07&7506&-144.26&142.35&2.59&2.59&202.67\\
\bottomrule[0.65pt] 
\end{tabular*}
\end{center}
\vspace*{-1mm} {\footnotesize \hspace*{10.5mm}a) {Proper motion is provided by PPMXL catalog}
\hspace*{10.5mm}b) {'NA' indicates proper motion is not available}
}
\end{table}

\begin{multicols}{2}
\begin{spacing}{1.05}

\begin{figure}[H]
\begin{center}
\includegraphics[width=0.45\textwidth]{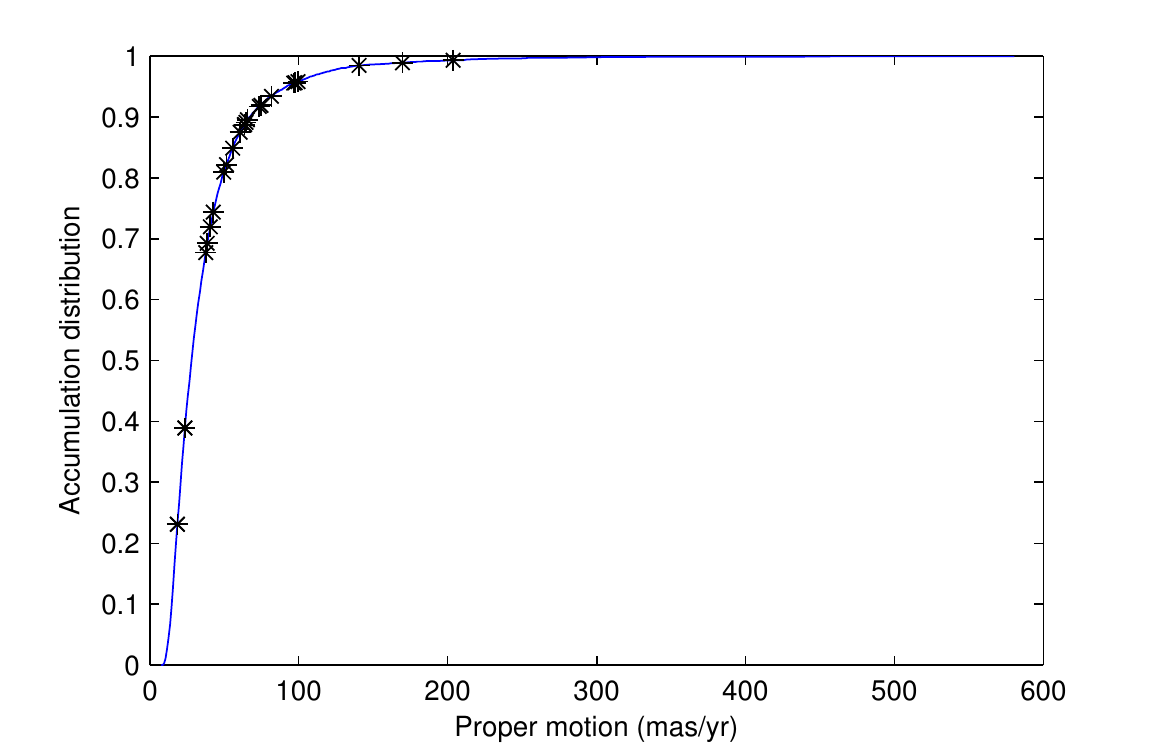} \\ 
\end{center}
\label{RL4}
\caption{Accumulation distribution of proper motions of 9374 white dwarfs identified by Kleinman [25] . The '*' are marked as proper motion of our 25 samples with high confident proper motions}
\end{figure}
\begin{table}[H]
\label{tab:table7}
\caption{Coefficients estimation}
\begin{center}\footnotesize \doublerulesep 0.2pt \tabcolsep 7pt
\begin{tabular}{*{4}l}
\hline
$a_3$& $a_2$ & $a_1$ & $a_0$\\
\hline
-0.1989&0.5428&-0.5335&3.9147\\
\bottomrule[0.65pt] 
\end{tabular}
\end{center}
\end{table}

\begin{figure}[H]
\begin{center}
\includegraphics[width=0.45\textwidth]{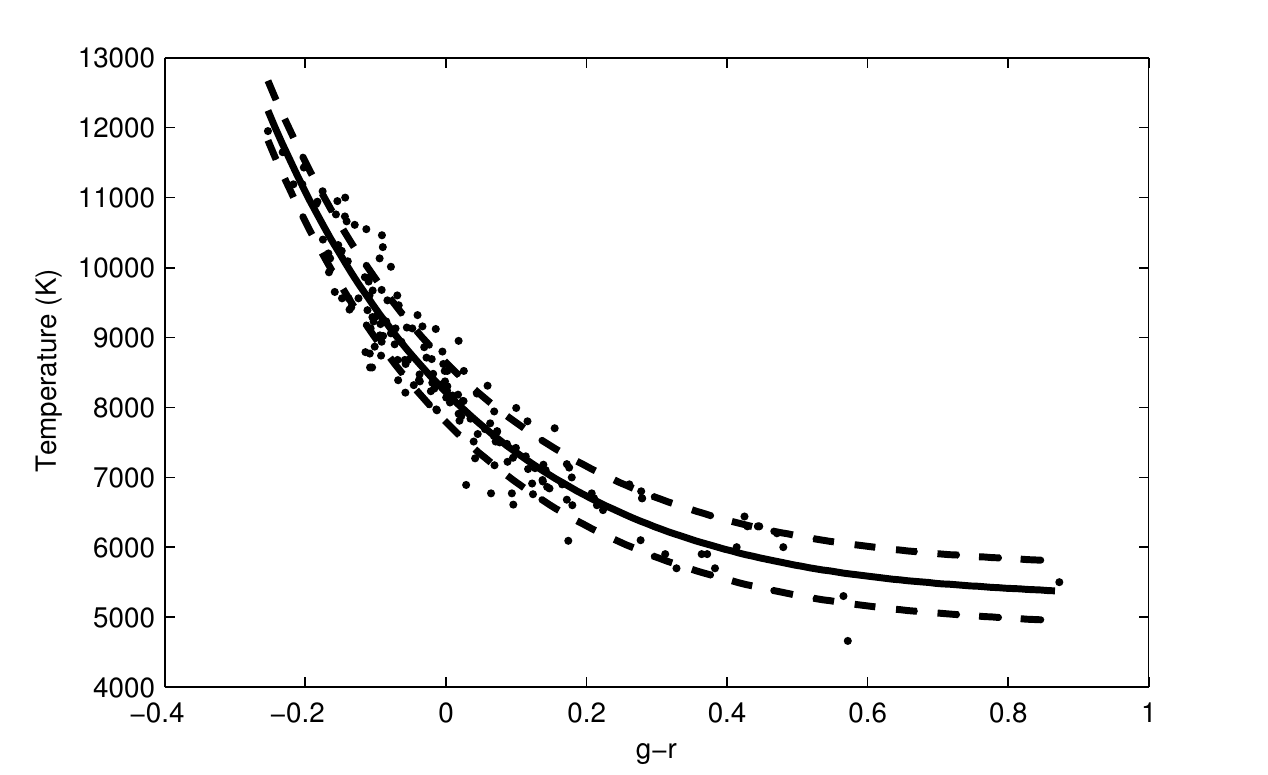} \\ 
\end{center}
\label{RL4}
\caption{Temperature fitting of DZ white dwarfs. The solid red curve is the fitting result and the two dotted curves indicate one $\sigma$. }
\end{figure}

\quad There were 173 DZ white dwarfs from SDSS whose temperatures were well determined, 147 of them were identified by Dufour et al. [23] two of which are below 6600K, and 26 of them were analysed by Koester et.al [24] which are extension of the well-known DZ sequence towards cooler temperatures, and fill the gap around 6500K. We obtained the relationship between their effective temperature and g-r color, which is shown in Eq.(7): 
\begin{equation}
log(T_{eff}/K) = a_3*(g-r)^3 + a_2*(g-r)^2 + a1*(g-r) + a_0 
\end{equation}
The coefficients can be estimated using least square fitting method, and are listed in Table 6. The fitting result is described in Fig. 5, the solid line is fitting curve and the two dotted lines indicate one standard deviation $\sigma = 425K$. Using Eq.(7), we estimated the effective temperatures  of our DZ white dwarfs, and found that nine of our samples are between 6000K and 6600K, and three of them are below 6000K. The spectra of these 12 cool DZ white dwarfs are plotted in Fig. 3.
\section{Conclusions}
Herein, we have firstly analysed the performance of the label propagation in searching for carbon stars from 656,801 stellar spectra of SDSS DR8, and concluded that the algorithm is efficient and robust to search for rare stars from massive spectra with only a few of known templates. However, the algorithm can still be improved. A key point, which is crucial for the performance, is how to construct proper graph. There are two elements to consider. One element is how to choose nice parameter $K$ and $\sigma$,  and the other one is how to select better features. In future, we will plan to focus on the two elements to improve the performance.

\quad Then, we have applied the label propagation algorithm to search for carbon stars from 656,801 star, 1,003,596 galaxy and 182,803 QSO spectra of SDSS DR8 respectively, and found 260 new carbon stars and 11 composition spectrum systems, each of which consists of a white dwarf and a carbon star. According to the criteria described in Green [22], there are 96 stars identified as dwarf carbon stars and 7 identified as giant carbon stars in our carbon stars. By manually checking our new carbon stars, we have found that 178 of them  are G- or F-type carbon stars, which are incorrectly classified as G- or F-type stars by SDSS pipeline. The reason why they are not classified correctly may be that they have similar spectra features with G- or F-type stars and the pipeline has no such type of carbon star templates. Our G- or F-type carbon stars and those identified by Green [22] should be sufficient to construct high-quality G- or F-type carbon star templates using method mentioned by Rosalie et al. [29]. In addition, the 11 composite spectrum are quite rare, and whether they are real physical binaries need second-epoch observation in the future.

\quad Finally, we also have applied the algorithm to search for DZ white dwarfs from all spectra of SDSS DR8, and found 29 new DZ white dwarfs. Their spectra were reconstructed by PCA algorithm with DZ white dwarf spectra given previously, and we have found that they are typical DZ white dwarfs through comparing with the reconstructed spectra. We have also investigated their proper motions, and found that they satisfy the fact that currently observed white dwarfs by SDSS generally have large proper motions. In addition, we have also estimated their effective temperatures by fitting the relationship between effective temperature and g-r color of DZ white dwarfs, and found 12 cool DZ white dwarfs, in which nine are between 6000K and 6600K, and three are below 6000K.

\end{spacing}
\Acknowledgements{\bahao We are grateful to Ren Juanjuan for the method of making cross identification to find proper motions. The work was funded by the National Science Foundation of China (Grant Nos. 10973021, 11303036). SDSS-III has been provided by the Alfred P. Sloan Foundation, the Participating Institutions, the National Science Foundation, and the U.S. Department of Energy Office of Science. The SDSS-III web site is http://www.sdss3.org/.
SDSS-III is managed by the Astrophysical Research Consortium for the Participating Institutions of the SDSS-III Collaboration including the University of Arizona, the Brazilian Participation Group, Brookhaven National Laboratory, Carnegie Mellon University, University of Florida, the French Participation Group, the German Participation Group, Harvard University, the Instituto de Astrofisica de Canarias, the Michigan State/Notre Dame/JINA Participation Group, Johns Hopkins University, Lawrence Berkeley National Laboratory, Max Planck Institute for Astrophysics, Max Planck Institute for Extraterrestrial Physics, New Mexico State University, New York University, Ohio State University, Pennsylvania State University, University of Portsmouth, Princeton University, the Spanish Participation Group, University of Tokyo, University of Utah, Vanderbilt University, University of Virginia, University of Washington, and Yale University. 
}


\normalsize \vskip0.1in\parskip=0mm \baselineskip 18pt
\renewcommand{\baselinestretch}{1.06}\footnotesize\parindent=4mm\bahao

\vskip0.1in \noindent 
\vskip0.1in\parskip=0mm

\REF{1\ }York D G, Adelman J, Anderson J E, et al. The Sloan Digital Sky Survey: technical summary. Astron J, 2000, 120: 1579--1587
\REF{2\ }Tu L P, Luo A L, Wu F C, et al. Reducing the searching range of Supernova candidates automatically in a flood of spectra of galaxies (in Chinese). Spectrosc Spect Anal, 2009, 29(12): 3420--3423 
\REF{3\ }Tu L P, Luo A L, Wu F C, et al. New Supernova candidates from the SDSS-DR7 spectral survey. Res Astron Astrophys, 2009, 9(6): 635--640
\REF{4\ }Tu L P, Luo A L, Wu F C, Zhao Y H. A method of searching for Supernova candidates from massive galaxy Spectra. Sci China Phys Mech Astron, 2010, 53(10): 1928-1938 
\REF{5\ }Wei P, Luo A L, Li Y B, Pan J C, et al. Mining unusual and rare stellar spectra from large spectroscopic survey data sets using the outlier-detection method. Mon Not R Astron Soc, 2013, 431(2): 1800--1811
\REF{6\ }Meusinger H, Schalldach P, Scholz R D, et al. Unusual quasars from the Sloan Digital Sky Survey selected by means of Kohonen self-organising maps. Astron Astrophys, 2012, 541: 77--104. 
\REF{7\ }Daniel S F, Connolly A, Schneider J, et al. Classification of stellar spectra with local linear embedding. Astron J, 2011, 142: 203--212
\REF{8\ }Shin M S, Yi H, Kim D W, et al. Detecting variability in massive astronomical time-series Data. II. variable candidates in the Northern Sky Variability Survey. Astron J, 2012, 143: 65--82 
\REF{9\ }Zhu X J, Ghahramani Z B. Learning from labeled and unlabeled data with label propagation. Technical Report CMU-CALD-02-107, 2002.
\REF{10\ }Wang F, Zhang C S. Label propagation through linear neighborhoods. In: Cohen W W, Moore A, eds. Proceedings of the 23rd International Conference on Machine learning. New York, USA, 2006. 985--992  
\REF{11\ }Cheng H, Liu Z C, Yang J. Sparsity induced similarity measure for label propagation. Proceedings of IEEE 12th International Conference on Computer Vision. Kyoto, Japan, 2009. 317--324 
\REF{12\ }Xu B, Bu J J, Chen C, et al. Efficient manifold ranking for image retrieval. Proceedings of the 34th international ACM SIGIR conference on Research and development in Information Retrieval. Beijing, China, 2011. 525--534 
\REF{13\ }Mould J R, Schneider D P, Gordon G A, et al. The velocity dispersion of carbon stars at the north Galactic pole. Publ Astron Soc Pac, 1985, 97: 130--137
\REF{14\ }Bothun G, Elias J H, MacAlpine G, et al. Carbon stars at high Galactic latitude. Astron J, 1991, 101: 2220--2228
\REF{15\ }Battinelli P, Demers S, Rossi C, et al. Extension of the C star rotation curve of the Milky Way to 24 kpc. Astrophysics, 2013, 56: 68--75.
\REF{16\ }Demers S, Battinelli P. C stars as kinematic probes of the Milky Way disk from 9 to 15 kpc. Astron Astrophys, 2007, 473: 143--148
\REF{17\ }Demers S, Battinelli P, Forest H. Outer MW kinematics from carbon stars. In: Andersen J, Bland-Hawthorn J, Nordstr\"{o}m B, eds. Proceedings of the International Astronomical Union, IAU Symposium, Cambridge: Cambridge University Press, 2009. 20--20
\REF{18\ }Aannestad P A, Kenyon S J, Hammond G L, et al. Cool metallic-line white dwarfs, radial velocities, and interstellar accretion. Astron J, 1993, 105(3): 1033--1044
\REF{19\ }Farihi J, Barstow M A,  Redfield S, et al. Rocky planetesimals as the origin of metals in DZ stars. Mon Not R Astron Soc, 2010, 404(4): 2123--2135 
\REF{20\ }Margon B, Anderson S F, Harris H C, et al. Faint high-latitude carbon stars discovered by the Sloan Digital Sky Survey: methods and initial results. Astron J, 2002, 124: 1651--1669
\REF{21\ }Downes R A, Margon B, Anderson S F, et al. Faint high-latitude carbon stars discovered by the Sloan Digital Sky Survey: an initial catalog. Astron J, 2004, 127: 2838--2849
\REF{22\ }Green P. Innocent Bystanders: Carbon stars from the Sloan Digital Sky Survey. Astrophys J, 2013, 765: 12--30
\REF{23\ }Dufour P, Bergeron P, Liebert J, et al. On the spectral evolution of cool, helium-atmosphere white dwarfs: detailed spectroscopic and photometric analysis of DZ stars. Astrophys J, 2007, 663: 1291-1308
\REF{24\ }Koester D, Girven J, G\"{a}nsicke B T, et al. Cool DZ white dwarfs in the SDSS. Astron Astrophys, 2011, 530: 114--124
\REF{25\ }Kleinman S J, Kepler S O, Koester D, et al. SDSS DR7 white dwarf catalog. Astrophys J Suppl, 2013, 204: 5
\REF{26\ }Heber U, Bade N, Jordan S, et al. PG 0824+289 - A dwarf carbon star with a visible white dwarf companion. Astron Astrophys, 1993, 267: L31--L43
\REF{27\ }Liebert G, Schmidt G D, Lesser M. Discovery of a dwarf carbon star with a white dwarf companion and of a highly magnetic degenerate star. Astrophys J, 1994, 421: 733--737
\REF{28\ }Munn, J A, Monet, D G, Levine, S E, et al. An improved proper-motion catalog combining USNO-B and the Sloan Digital Sky Survey. Astron J, 2004, 127: 3034--3042
\REF{29\ }McGurk R C, Kimball A E, Ivezi\'{c} Z. Principal component analysis of Sloan Digital Sky Survey stellar spectra. Astron J, 2010, 139: 1261--1268

\end{multicols}

\end{document}